\documentclass[conference]{IEEEtran}
\IEEEoverridecommandlockouts
\usepackage{tabularray}
\usepackage{cite}
\usepackage{amsmath,amssymb,amsfonts}
\usepackage{algorithmic}
\usepackage{graphicx}
\usepackage{textcomp}
\usepackage{xcolor}
\def\BibTeX{{\rm B\kern-.05em{\sc i\kern-.025em b}\kern-.08em
    T\kern-.1667em\lower.7ex\hbox{E}\kern-.125emX}}
\begin{document}

\title{AutoTest: Evolutionary Code Solution Selection with Test Cases }

\author{\IEEEauthorblockN{1\textsuperscript{st} Zhihua Duan}
\IEEEauthorblockA{\textit{Intelligent Cloud Network Monitoring Department} \\
\textit{China Telecom Shanghai Company}\\
\textit{700 Daning Road, Shanghai, 200072}\\
Shanghai,China \\
duanzh.sh@chinatelecom.cn}
\and
\IEEEauthorblockN{2\textsuperscript{nd} Jialin Wang}
\IEEEauthorblockA{\textit{Computer Science} \\
\textit{Stanford University}\\
\textit{450 Serra Mall, Palo Alto,94305}\\
California, America \\
jialinwangspace@gmail.com}
 
}

\maketitle

\begin{abstract}
With the development of code generation techniques, selecting the correct code solution from multiple candidate solutions has become a crucial task. This study proposes AutoTest, a novel technique that combines automated test case generation with code solution execution to optimize the selection process using an evolutionary genetic algorithm. Firstly, AutoTest utilizes large pre-trained language models such as codegen-16B, code-davinci-002, and incoder-6B to provide code solutions and their corresponding test cases. Then, by executing the code solutions and evaluating their performance on the test cases, a consensus set is formed. Fine-grained ranking is achieved through the selection, mutation, and crossover mechanisms based on the evolutionary genetic algorithm, with the adjustment of alpha and beta parameters. Finally, the best code solution is chosen.

AutoTest demonstrates significant performance improvements on the HumanEval benchmark test. The HumanEval dataset consists of 164 programming problems, and AutoTest achieves approximately a 10\% improvement over the baseline method in terms of pass@1 score. 
\end{abstract}


\section{Introduction}
Code generation technology has made significant progress in recent years, but selecting the correct solution from multiple candidate solutions remains a challenge\cite{LLM-TEST-AUTOMATED,LLM-TEST-CODET}. This paper proposes the AutoTest method, which combines the steps of automatically generating test cases and executing code solutions to select the correct solution using an evolutionary genetic algorithm.

As shown in Figure 1, the first step involves reusing large models such as codegen-16B\cite{TTM-TEST-codegen-16B}, code-davinci-002\cite{LLM-TEST-code-davinci-002}, and incoder-6B\cite{LLM-TEST-INcoder} to generate code solutions and test cases for each programming problem. Next, each solution is executed on the generated test cases, and a consensus set is formed by finding commonalities among multiple solutions and test cases. Then, by adjusting the alpha and beta parameters and applying the selection, mutation, and transformation mechanisms of evolutionary genetics, the code solutions are finely ranked, and the highest-ranked code solution is selected as the best solution.
\begin{figure}[h]
  \centering
  \includegraphics[width=\linewidth]{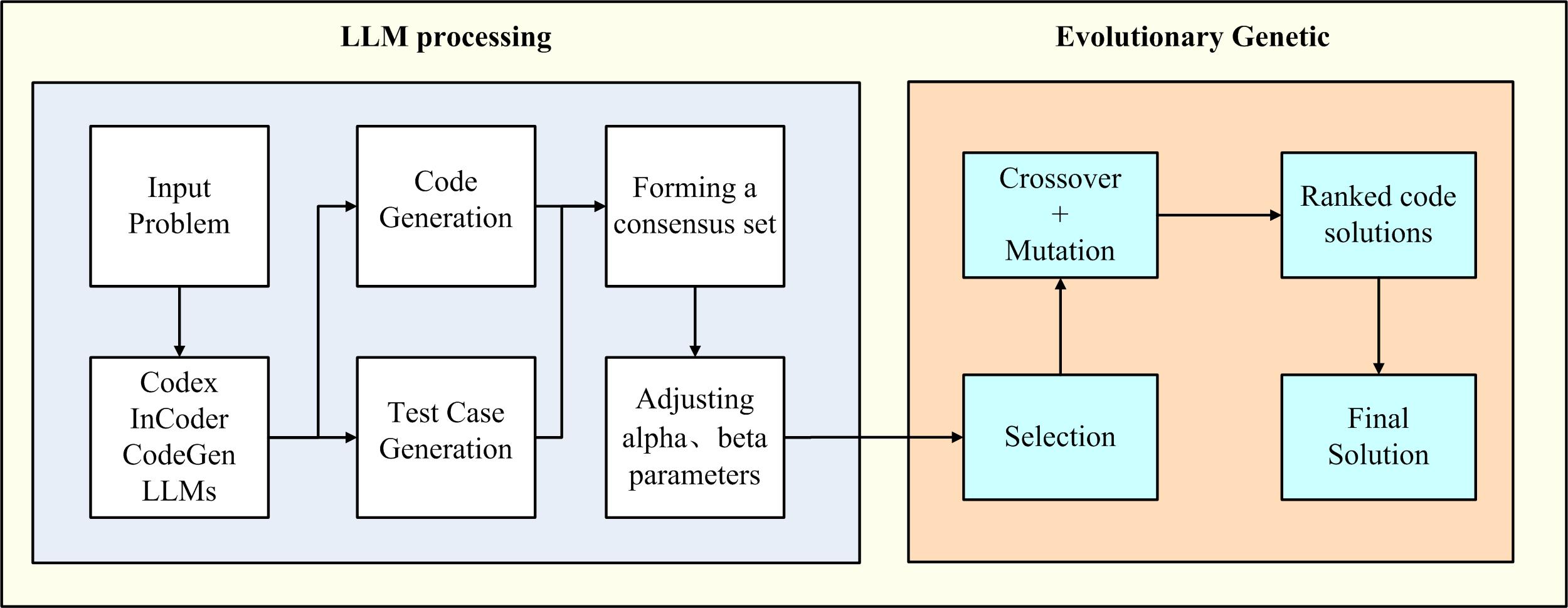}
  \caption{The illustration of AutoTest}
   
\end{figure}
The AutoTest method is simple and efficient. Experimental results demonstrate that AutoTest achieves significant improvements on the HumanEval benchmark test. By combining the automatic generation of test cases with the execution of code solutions, the AutoTest method is able to select the correct code solution from multiple candidate solutions, thereby improving the pass@1 score.
 
\section{Related Work}
Large language models leverage abundant open-source code resources and undergo training with billions of parameters, showcasing remarkable performance in code generation tasks. For instance, models such as AlphaCode\cite{LLM-TEST-Alphacode}, Codex\cite{LLM-TEST-OpenAI-HumanEval}, CodeT5\cite{LLM-TEST-CODET5}, CodeGen\cite{TTM-TEST-codegen-16B}, InCoder\cite{LLM-TEST-INcoder}, CodeGeeX\cite{LLM-TEST-CodeGeeX} and StarCoder\cite{LLM-TEST-StarCoder2} have achieved significant advancements in code generation applications. 

By utilizing automated test case generation techniques, it is possible to greatly enhance work efficiency. Pretrained language models such as BART, T5, Codex, and other large-scale models can be employed in a zero-shot setting to sample and generate test cases using prompt statements.

Despite the excellent performance of large-scale models in code generation tasks, they still require iterative sampling to determine the correct output. Researchers such as \cite{LLM-TEST-NATURAL}, and others have proposed solutions to address this issue. In contrast to their work, this paper leverages the code solutions and test cases provided by large-scale models and employs an evolutionary genetic algorithm to select the best code solution.

\section{METHODOLOGY}
The goal of code generation tasks is to generate code solutions based on contextual prompts, which include natural language problem descriptions and code snippets. Using large language models, a set of code solutions is generated based on the contextual prompts, and the best code solution is selected from the generated set. The AutoTest method utilizes pre-trained language models to provide code solutions and test cases related to code programming problems.
\begin{itemize}
\item Code Solutions: Powerful large-scale language models such as codegen-16B, code-davinci-002, and incoder-6B can be employed to generate code solutions. These models possess extensive programming knowledge and capabilities, allowing them to generate high-quality code based on the provided requirements and contextual information.

\item Test Cases: Test cases are generated to evaluate the correctness of the code solutions. Using the same pre-trained language model used for generating code solutions, input and expected output for defining functions are specified within the context to generate test cases. It is ensured that the generated test cases exhibit diversity and difficulty.
\end{itemize}
In this paper, an evolutionary genetic algorithm is employed to select the best code solution. The fundamental idea behind the evolutionary genetic algorithm is to execute the generated code solutions and compare their execution results on the generated test cases to determine their correctness. Through iterative execution and comparison, consistent combinations of code solutions and test cases are identified. The evolutionary genetic algorithm is then utilized to select the highest-ranked solution based on the included test cases and solutions as the best code solution.

As shown in the figure 2, AutoTest proposes a method for selecting the best code solution based on the provided set of code solutions and test cases. By executing the code solutions and comparing them with the expected outputs of the test cases, it can be determined whether the code solutions pass the tests.

\begin{figure}[h]
  \centering
  \includegraphics[width=\linewidth]{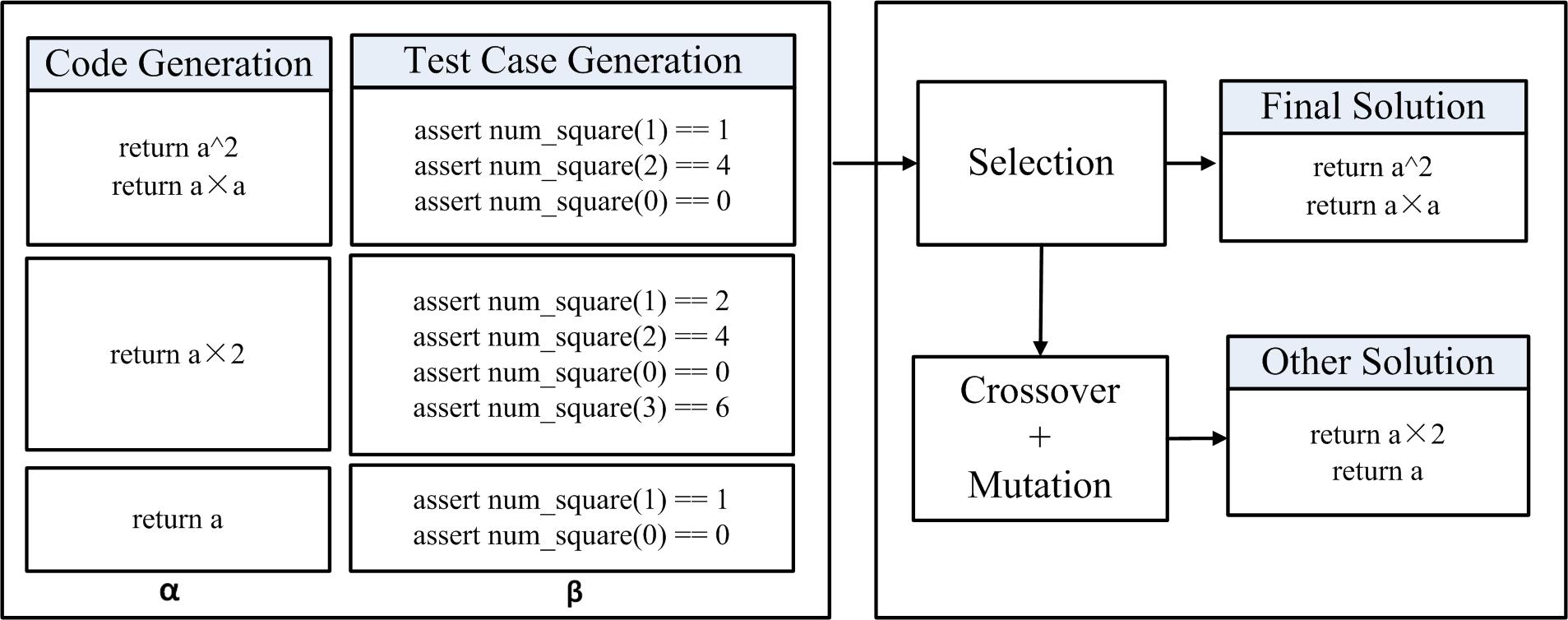}
  \caption{A simple example: Obtaining the code solution for "return the square of a number."}
   
\end{figure}

The AutoTest method follows the following steps for implementation:

1. Randomly select a pair of code solutions and test cases and attempt to execute the code solution on the test cases. If the code solution passes the test cases, it is labeled as an inlier; otherwise, it is labeled as an outlier. For each inlier, collect other code solutions and test cases that are consistent with it to form a consensus set.

For example, both "return a\^{}2" and "return a×a" pass the test cases "assert num\_square(1) == 1," "assert num\_square(2) == 4," and "assert num\_square(0) == 0," forming a collection of code solutions and test cases with 2 code solutions and 3 test cases.

2. Based on the consensus set of code solution and test case pairs,Assuming alpha is set to 0.5 and beta is set to 1.1,use an evolutionary genetic algorithm with selection, mutation, and crossover mechanisms to rank the code solutions. Individuals with higher fitness are selected as parents, and mutation and crossover operations introduce new solutions and generate offspring. 

3. Select the code solution with the highest score from the consensus set as the best solution. the best code solution obtained through the evolutionary genetic algorithm is either "return a\^{}2" or "return a×a."

AutoTest proposes an automated testing method based on code generation solutions and test cases from large language models. By utilizing code solutions and test cases generated by the large language model, the method employs an evolutionary genetic algorithm to select the best code solution.

The alpha parameter controls the weight assigned to code generation, and the beta parameter controls the weight assigned to test case generation. By adjusting the values of alpha and beta, AutoTest can adapt to different coding problems and datasets. The choice of optimal values for alpha and beta depends on the characteristics of the problem domain, the complexity of the code solutions, and the nature of the test cases.

AutoTest algorithm allows for exploration and evaluation of diverse code solutions and test cases, reducing the dependence on any single model and mitigating the impact of biases and overfitting. 

\section{EXPERIMENT}
\subsection{Dataset}
The HumanEval\cite{LLM-TEST-OpenAI-HumanEval} dataset is a collection of 164 programming problems designed for evaluating the correctness of functions. Each problem consists of a function signature, a documentation string, a function body, and multiple unit tests, with an average of 7.7 tests per problem. As the large language models are trained on code from GitHub, which contains solutions from various sources, the programming tasks in the HumanEval dataset are manually crafted to assess language comprehension, reasoning abilities, algorithms, and basic mathematical skills. The dataset serves as a measure of the problem-solving capabilities of the large language models.
\subsection{Models}
This study utilized experimental data provided by Microsoft \cite{LLM-TEST-CODET}, which tested the codegen-16B, code-davinci-002, and incoder-6B large-scale language models. The code-davinci-002 model, developed by OpenAI, is a large language model optimized specifically for code generation tasks. It can be applied in various scenarios such as code autocompletion and code inference. The incoder-6B model, released by Meta, focuses on code understanding and generation. The codegen-16B model, proposed by Salesforce, is a large language model with the ability to generate or modify code based on natural language prompts.

\subsection{Metrics}
The pass@k (n samples) metric is used for performance evaluation, and test cases are employed to verify the correctness of the code solutions. For each problem, n code solutions are randomly selected from the samples, and k solutions are chosen for evaluation. If any of these selected solutions pass all the test cases, it is considered that the problem has been successfully solved. The pass@k metric represents the percentage of problems solved. An unbiased definition of pass@k, as a benchmark, involves randomly selecting k solutions from n samples. In this study, an evolutionary genetic next-generation algorithm protocol is applied. It selects k solutions from n samples and dynamically adjusts alpha and beta values. Methods such as selection, alteration, and mutation are used to group the solutions based on test outputs and sort them according to their scores.

\subsection{Experimental Results}  
The evaluation of the HumanEval benchmark test set was conducted based on the Codex, InCoder and CodeGen pretrained models. The experimental results on the HumanEval benchmark test for the codegen-16B, code-davinci-002, and incoder-6B large models are summarized in Table I. In the baseline methods, the pass@100 scores of the large models were significantly higher than the pass@1 scores, indicating a significant advantage in selecting the best code solution from 100 samples. Additionally, in the AlphaCode and AutoTest methods, pass@2 and pass@10 scores were also evaluated in this study. Comparing the AutoTest scores with the baseline scores for pass@1, it was found that the AutoTest models achieved an improvement of approximately 10\% compared to the baseline pass@1. For pass@2, AutoTest showed an improvement over the AlphaCode scores. For pass@10, the AutoTest score for code-davinci-002 was 85.0, which is a 0.6\% improvement over the AlphaCode score of 84.4. These results demonstrate that AutoTest can enhance the performance of various pretrained language models.

\begin{table}
\caption{Pass@k (\%) on the HumanEval benchmarks,Scores with  $^{\dag}$ were obtained from \cite{LLM-TEST-CODET}} 
\centering
\begin{tblr}{
  width = \linewidth,
  colspec = {Q[187]Q[285]Q[135]Q[135]Q[173]},
  cell{2}{1} = {r=3}{},
  cell{6}{1} = {r=3}{},
  cell{10}{1} = {r=3}{},
  hline{1-2,5,9,13} = {-}{},
  hline{3-4,6-8,10-12} = {2-5}{}, 
}
Method    & Model            & pass@1 & pass@2 & pass@100 \\
Baseline{$^{\dag}$}  & codegen-16B      & 29.7   & 50.3   & 73.7     \\
          & code-davinci-002 & 47.0     & 74.9   & 92.1     \\
          & incoder-6B       & 16.4   & 28.3   & 47.5     \\
          & Model            & pass@1 & pass@2 & pass@10  \\
AlphaCode{$^{\dag}$}  & codegen-16B      & 27.3   & 38.5   & 64.4     \\
          & code-davinci-002 & 55.1   & 64.1   & 84.4     \\
          & incoder-6B       & 17.7   & 23.8   & 34.8     \\
          & Model            & pass@1 & pass@2 & pass@10  \\
AutoTest  & codegen-16B      & 36.8   & 44.1   & 58.2     \\
          & code-davinci-002 & 64.5   & 74.5   & 85.0     \\
          & incoder-6B       & 20.1   & 25.9   & 37.0     
\end{tblr} 
\end{table}

\subsection{Discuss}
This paper uses an evolutionary genetic algorithm for optimizing solution approaches. Further research is needed to evaluate its effectiveness in scoring and solving code-related problems. Nonetheless, the study offers promising directions for future research.

1. Improved scoring mechanism: Revisiting the scoring mechanism, the original scoring method may have certain limitations. To address this issue, new scoring metrics can be introduced to better measure the performance of solutions.

2. Optimization of algorithm parameters: Optimizing and adjusting the alpha and beta parameters of the evolutionary genetic algorithm. 

Future research can further experiment and compare these methods to validate the effectiveness and superiority of the approach, providing valuable directions for further study and application.

\section{CONCLUSION}
This paper proposes a large language model based approach called AutoTest, which utilizes a large model to generate code solutions and test cases simultaneously. The generated test cases are then executed to evaluate the code solutions, and an evolutionary genetic algorithm is employed to select the best solution. Future work can focus on further improving the AutoTest method to address more complex programming problems, support multiple programming languages, handle more sophisticated algorithms, and enhance the diversity and coverage of test case generation.

\bibliographystyle{unsrt}

\end{document}